\documentclass[]{aa}
\usepackage{graphics,epsf}
\begin{document}

\def\ve{{\vec e}} \def\vl{{\vec l}} \def\vx{{\vec x}}  \def\vk{{\vec
k}} \def\vgam{{\vec{\gamma}}} \def\mD{{\cal{D}}}  \def\d{{\rm d}}
\def\i{{\rm i}} \def\veps{{\vec \epsilon}} \def\eps{\epsilon}
\def\gam{\gamma}

\titlerunning{Non-Gaussian Signatures in Lensing Survey}
\title{Detection of non-Gaussian Signatures in the VIRMOS-DESCART
Lensing Survey\thanks{Based on observations obtained at the
Canada-France-Hawaii Telescope (CFHT) which is operated by the
National Research Council of Canada (NRCC), the Institut des Sciences
de l'Univers (INSU) of the Centre National de la Recherche
Scientifique (CNRS) and the University of Hawaii (UH).}}

\author{Francis Bernardeau\inst{1} \and Yannick Mellier\inst{2,3} \and
Ludovic van Waerbeke\inst{2,4}} \institute{Service de Physique
Th{\'e}orique, CE de Saclay 91191 Gif-sur-Yvette Cedex, France \and
Institut d'Astrophysique de Paris, 98 bis, boulevard Arago, 75014
Paris, France \and Observatoire de Paris, DEMIRM/LERMA, 61 avenue de
l'Observatoire, 75014 Paris, France  \and Canadian Institute for
Theoretical Astrophysics, 60 St George St., Toronto, Ont. M5S 3H8,
Canada}

\newcommand{\etal}{{\sl et al. }}  \newcommand{\beq}{\begin{equation}}
\newcommand{\eeq}{\end{equation}}

\date{Received 27 February 2002 / Accepted 14 May 2002}

\abstract{We have  detected non-Gaussian signatures in the
VIRMOS-DESCART weak lensing survey from a measurement of the
three-point shear correlation function, following the method
developped by Bernardeau, van Waerbeke and Mellier (2002). We obtain a
2.4$\sigma$ signal over four independent angular bins, or
equivalently, a 4.9-$\sigma$ confidence level detection with respect
to measurements errors on scale of about $2$ to $4$ arc-minutes. Both
amplitude and shape are found to agree with theoretical expectations
that have been investigated for three cosmological models. This supports
the idea that the measure corresponds to a  cosmological signal due to
the gravitational instability dynamics. Its properties could be used
to put constraints on the cosmological parameters, in particular on
the density parameter of the Universe, but the error level as well as
the cosmic variance are  still too large to permit secure conclusions.
\keywords{cosmology, gravitational lensing, large scale structure}}

\maketitle

\section{Introduction}

The large-scale structure of the Universe are expected to form from
the gravitational growth of initial density perturbations obeying
Gaussian statistics. As the Universe expands and the perturbations
grow, non-Gaussian features are expected to emerge in the density
field due to gravitational dynamics.
 These features can be  characterized with perturbation theory
calculations, which allow to compute for instance the skewness, third
moment of the local density probability distribution function (Peebles
1980, Fry 1984, Bernardeau 1992). The reduced skewness of the density
field has been found to be quite insensitive to the variance and the
cosmological parameters, $\Omega_m$ (Juszkiewicz et al. 1992,
Bernardeau 1994). In contrast,  weak lensing surveys   are sensitive
to $\Omega_m$ since they trace the integrated mass along the
line-of-sight which is roughly proportional to the density parameter
of the Universe. Weak lensing by large scale structures has been
measured by several teams as a coherent distortion field of distant galaxies
over large angular distances (Bacon et al. 2000 and 2002;
H{\"a}mmerle et al.; 2002; Hoekstra et al. 2002; Kaiser et al. 2000;
Maoli et al. 2001; R{\'e}fr{\'e}gier et al 2002;  Rhodes et al. 2001; Van
Waerbeke et al. 2000, 2001 and 2002; Wittman et al. 2000). The projected
mass density reconstructed from the distortion field ($i.e.$ the
convergence field) can be used for non-Gaussian signatures searches,
as shown in Bernardeau, van Waerbeke \& Mellier (1997, hereafter
BvWM97). This work and further studies (Jain  \& Seljak 1997; van Waerbeke,
Bernardeau \& Mellier 1999) have
shown that the non-Gaussian properties of the convergence field can be
used as a probe of the cosmological density parameter, with a weak
dependence on the cosmological constant $\Omega_{\Lambda}$, provided
that the redshift distribution of the sources is known.

However, a straight  application of these theoretical considerations
to real data sets turned out to be arduous.  The convergence
field has to be recovered from a mass reconstruction process which
uses a continuous shear field obtained from a smoothed map of the
discrete galaxy ellipticities. Unfortunately, survey topologies are
 generally complex and are alterated by masked areas due to light scattering,
bright stars, comet-like reflections, asteroid/airplane tracks, very
bright galaxies, etc.... Since the mask sizes cover a range of scales
from a few arc-seconds to two degrees and are strongly anisotropic
(for instance bright stars preferentially saturate along CCD columns),
 results of mass reconstruction in such data sets are not yet reliable. An
alternative approach is the aperture mass applied to cosmic shear
(Schneider et al. 1998), which allows the measurement of the skewness
from the distortion field directly, bypassing the mass reconstruction
process. So far, our attempts for measuring the skewness of the
aperture mass lead to very noisy and un-significant results. Recently,
Bernardeau, van Waerbeke \& Mellier (2002, hereafter BvWM02)  have
proposed a new method using some specific patterns in the shear
three-point function. This method has also the advantage to bypass
the mass reconstruction process. Despite the complicated shape of the
three-point correlation pattern, BvWM02 uncovered a specific
geometrical property and demonstrated it can be used for the measurement
of non-Gaussian
 features. Their  detection strategy based on this method has
been found to be robust, usable in patchy catalogs, and quite
insensitive to the topology of the survey.  In the following we apply
this method to the VIRMOS-DESCART weak lensing survey done at the
Canada-France-Hawaii Telescope.

\section{Optimized 3-point correlation function applied to VIRMOS-DESCART
data} \label{sec:corr}
\subsection{Method}
Let us consider a triplet of galaxies at locations $\vx_1$, $\vx_2$
and $\vx'$, and their shear estimates $\vec\gamma(\vx_1)$,
$\vec\gamma(\vx_2)$ and $\vgam(\vx')$. BvWM02 introduced a 2D angular
vector field (representing 2 components of the general shear 3-point
correlation functions),
\begin{equation}
\xi_3(\vx')=\langle(\vec\gamma(\vx_1)\cdot\vec\gamma(\vx_2))\vgam(\vx')\rangle
\end{equation}
viewed as a function of $\vx'$  for a fixed
$\vert\vx_2\!-\!\vx_1\vert$ distance. At fixed separation $\vx_{12}$,
the components of $\xi_3$ are expected to scale like the square of the
shear 2-point correlation function,
$\xi_2(\vx_{12})=\langle\vec\gamma(\vx_1)\cdot\vec\gamma(\vx_2)\rangle$.
Its sensitivity to the cosmological density parameter ($\Omega$) is
expected to be similar to the convergence skewness ($\approx
\Omega^{-0.8}$), with an additional dependence on the slope of the
mass power spectrum. The way $\xi_3(\vx')$ varies as a function of
$\vx'$ is in general complicated although asymptotic properties can be
obtained analytically. However $\xi_3(\vx')$ has been found to be
rather uniform, and perpendicular to $\vx_{12}$, over an elliptic area
that covers the segment joining $\vx_1$ to $\vx_2$. This central
pattern turns out  to be robust against different cosmologies and
smoothing scales, with an amplitude which can be related to the
cosmological parameters. The results obtained by BvWM02 in synthetic
catalogs suggested that present-day cosmic shear surveys were already
large for a secure detection.


We consider the geometrical average,
\begin{eqnarray}
\overline{\xi_3}(\vert\vx_1-\vx_2\vert)= \int_{\rm Ell.}{\d^2\vx'\over
V_{\rm Ell.}} \xi_3(\vx') \label{xi3_ellip}
\end{eqnarray}
which corresponds to the average three-point function inside an
elliptic area, $V_{\rm Ell}$, encompassing the points $\vx_1$ and
$\vx_2$ (they are actually the foci of the ellipse). In this domain
$\xi_3(\vx')$ is expected to vary weakly and its average value
$\overline{\xi_3}$ is a vector quantity orthogonal to the $\vx_{12}$
direction\footnote{mathematically it means that the only non-zero
component of $\overline{\xi_3}$ is along $\vx_1-\vx_2$, but has a negative value.}
that depends only on the $\vx_1$-$\vx_2$ pair separation. An optimum
selection of pair points, where close pairs and highly elliptical
galaxies are rejected, turns out to provide  an optimal information on
non-Gaussian features in simulated catalogues.  BvWM02 checked that
different sources of noise produced by the intrinsic ellipticity
distribution of galaxies, by realistic galaxy shape measurements and
PSF corrections and  by masking effects do not spoil the result.
Final tests and validations  were carried out with simulated
catalogues containing as many galaxies as real data, with ellipticity
distribution, PSF anisotropy and masking templates similar to the
VIRMOS-DESCART sample. In all configurations the global signal to
noise remains higher than 5 for scales between 30 arc-seconds to 5
arc-minutes.

\subsection{The VIRMOS-DESCART sample}
The VIRMOS-DESCART sample used in this work is part of the DESCART
cosmic shear programme\footnote{http://terapix.iap.fr/Descart} which
uses the VIRMOS photometric/imaging
survey\footnote{http://www.astrsp-mrs.fr} for wide field cosmic
shear. It  covers 11.5 deg$^2$ of CFH12K images spread over four
uncorrelated fields. All data were   obtained in I-band up to a
limiting magnitude of $I_{AB}=24.5$ (within  5 arc-second aperture,
5-$\sigma$), which is consistent with a mean source redshift of
$z\simeq 0.9$ of the VIRMOS-DESCART sample (van Waerbeke et al. 2002).
The data contain all   observations used by van Waerbeke et al. (2001)
plus new fields  obtained  in September and November 2000.  All the
images were processed as described  in    van Waerbeke et al (2001) at
the TERAPIX data center\footnote{http://terapix.iap.fr}.  From an
initial detection we build up a  sample containing  1.6 millions
objects. After masking  and all galaxy selection processes,  it
reduces to 580,000  galaxies, covering an effective area of 8.5
deg$^2$.  Close-pairs with angular separation smaller than 10
arc-seconds can produce systematics and  are  rejected (see van
Waerbeke et al. 2000). \\
Following standard notation, the shear is defined as the mean source
ellipticities
\begin{equation}
\gamma_1=\langle e \cos(2\theta)\rangle, \ \ \ \gamma_2=\langle e
\sin(2\theta)\rangle \label{eqn:gamma}
\end{equation}
where $\theta$ is the angle between the major axis of the source
galaxy and the $x$ axis, and $e=(a-b)/(a+b)$ is determined by the
major axis length $a$ and minor axis length $b$.  It is computed
according to  the rules and weighting schemes given in Pen,  van
Waerbeke \& Mellier (2002) and the PSF anisotropy
  corrections used in van Waerbeke et al. (2002). In particular, we consider in this {\sl
Letter} the decomposition into $E$ and $B$ modes described in Pen et
al.  (2002) which is used for residual systematics checks for the
2-point statistics.

The statistical estimators for the binned two and three point
functions are given by
\begin{eqnarray}
\label{finalproduct} \xi_2(d_{ij})&=&{\sum_{ij}
w_i\,w_j\,(\ve_i\cdot\ve_j) \over \sum_{ij} w_i\,w_j}\\
 \overline{\xi}^{\ t}_3(d_{ij})&=& {\sum_{ijk}
 w_i\,w_j\,w_k\,(\ve_i\cdot\ve_j)\ e^{(ij)}_k \over \sum_{ijk}
 w_i\,w_j\,w_k }
\end{eqnarray}
where $e^{(ij)}_k$ is opposite to the  component of the ellipticity of
galaxy $k$ along the ($\vx_j-\vx_i$)-direction.  The summations are
made for pairs or triplets such that $d_{ij}=\vert\vx_i-\vx_j\vert$ is
in the chosen bin.\ \  $\vx_k$ lies within the ellipse defined by
$\vert\vx_k\!-\!\vx_i\vert +
\vert\vx_k\!-\!\vx_j\vert>1.1\vert\vx_i\!-\!\vx_j\vert$ and $w_i$ are
weights associated to each galaxy according to  the scheme discussed
in  BvWM02.

\subsection{The VIRMOS-DESCART 2 and 3-point correlation functions}
\begin{figure}
\begin{tabular}{c}
{\epsfxsize=7.5cm\epsffile{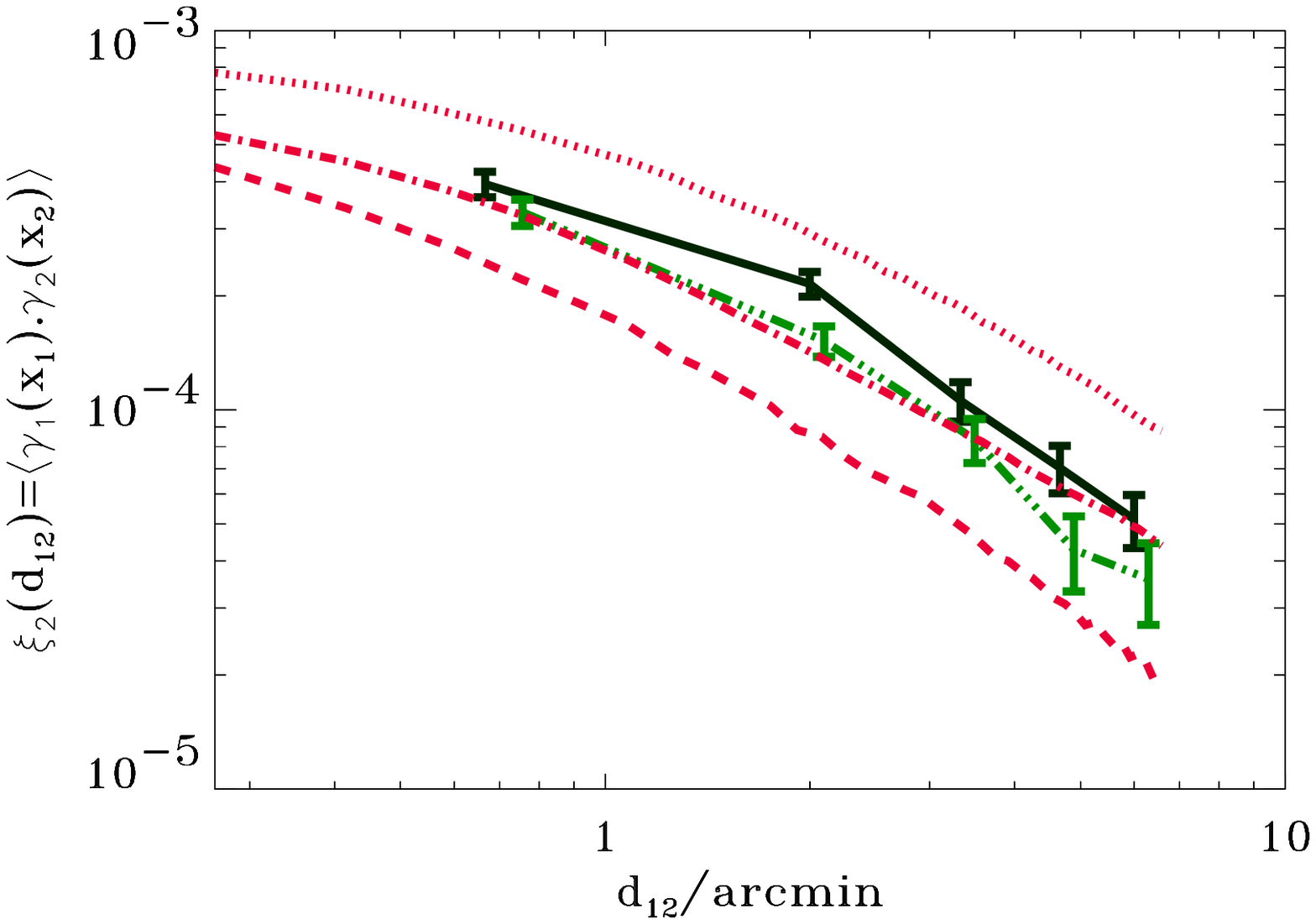}}\\
{\epsfxsize=7.5cm\epsffile{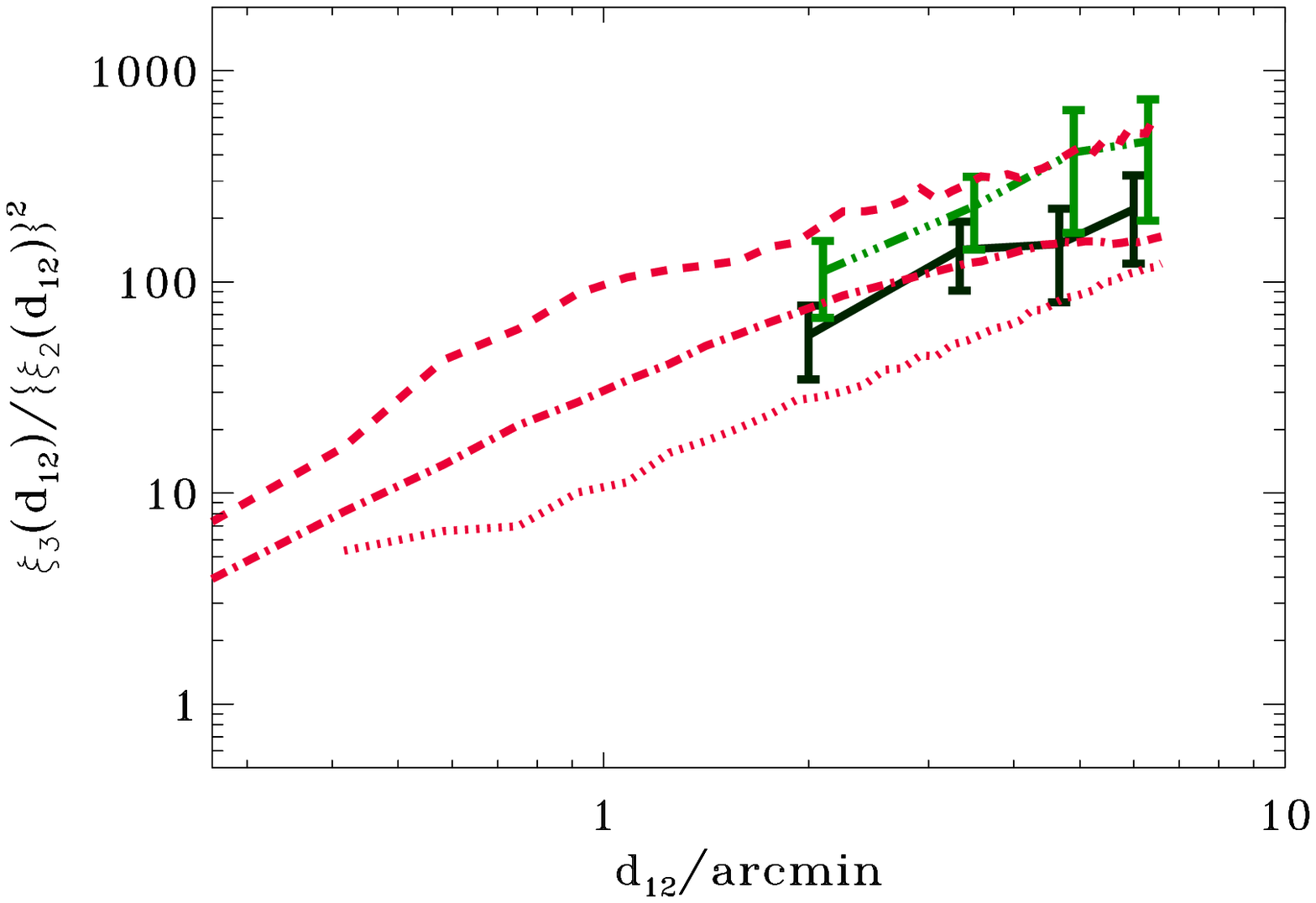}}
\end{tabular}
\caption{Results for the VIRMOS-DESCART survey
 for the two point correlation function
(top) and the reduced three point function (bottom).  The solid line
 with error bars shows the raw results, when both the $E$ and $B$ contributions
to the two-point correlation functions are included.  The  dot-dashed
line with error bars corresponds to measurements where the contribution of the $B$
mode has been substracted out from the two-point correlation function
(but not from $\xi_3$ there is no known way to do it). These
measurements are compared to results obtained in $\tau $CDM, OCDM and
$\Lambda$CDM simulations (dashed, dotted and dot-dashed lines
respectively). ($\tau $CDM, OCDM simulations are described in Jain,
Seljak \& White 2000; $\Lambda$CDM simulation was provided by
M. White.)} \label{xidetec}
\end{figure}

\begin{figure}
\begin{tabular}{c}
{\epsfxsize=7.5cm\epsffile{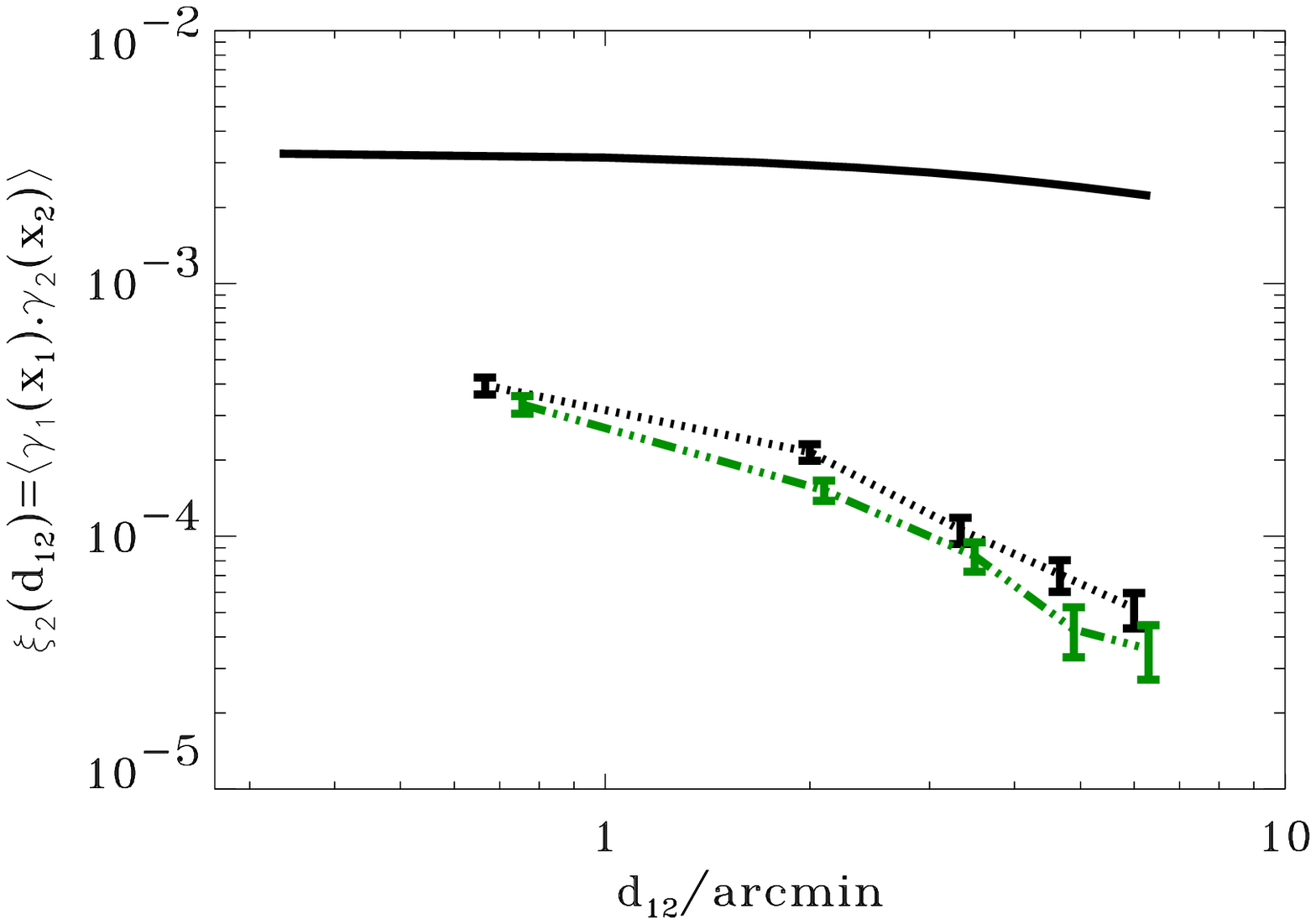}}\\
{\epsfxsize=7.5cm\epsffile{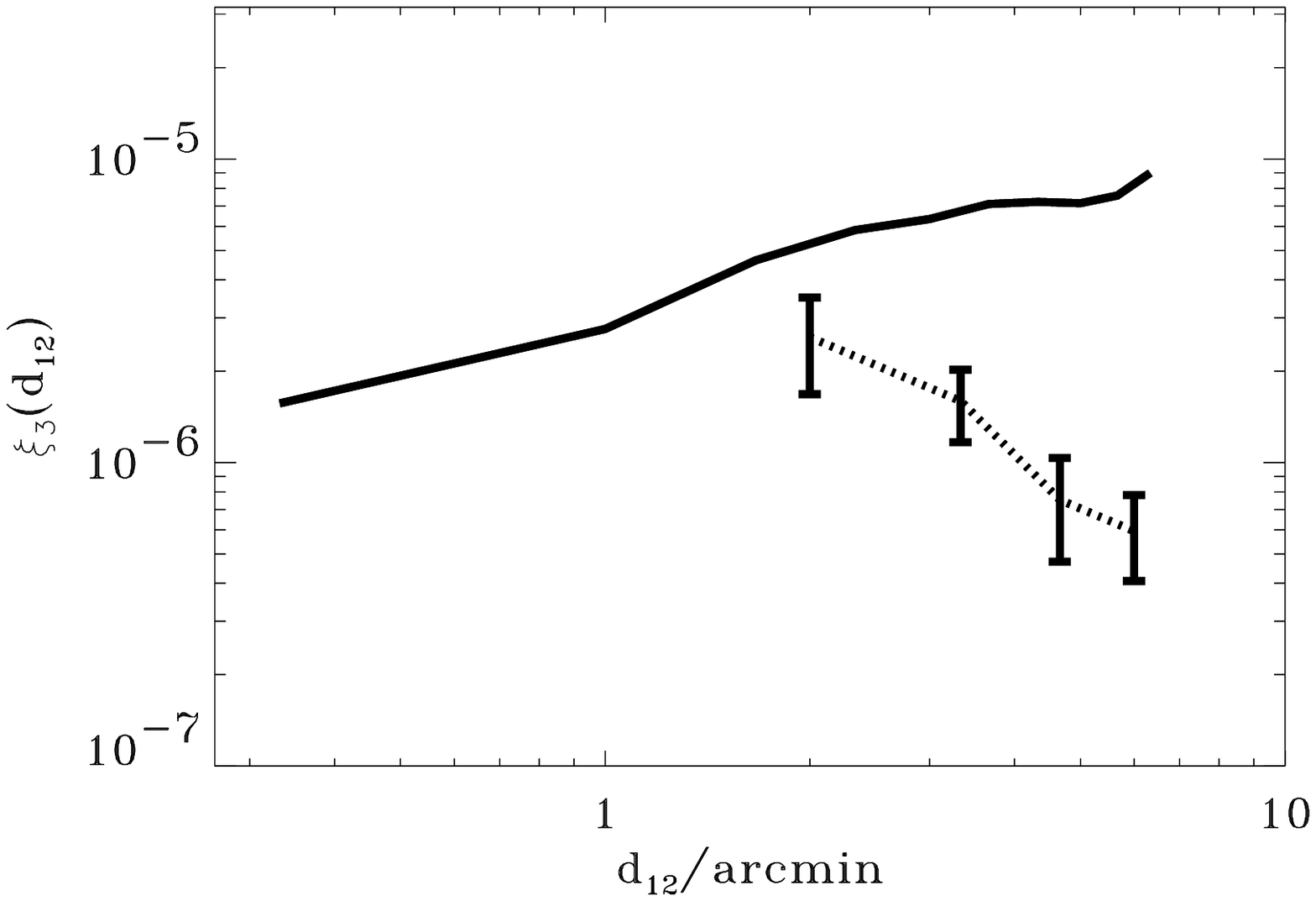}}
\end{tabular}
\caption{Results for the VIRMOS-DESCART survey (dashed lines:
when $E$ and $B$ mode contributions are included, dot-dashed line when
$B$ mode contributions are substracted out from the 2-point
ocrrelation function) compared the properties
of the star anisotropy field statistics (solid lines).}
\label{xistars}
\end{figure}

Figure \ref{xidetec} shows the estimated 2-point (top) and reduced
3-point (bottom) correlation functions ($\xi_3$ in units of
$\xi_2^2$). Measurements have been made in regularly spaced bins of
width 400 pixels (e.g. 1.3 arcmin). For comparison,  the thick dotted
lines shows the corresponding quantity for an open CDM model, the thick
dashed line for a $\tau $CDM and the dot-dasled line for a
$\Lambda$CDM model. The source redshift is one, very close to our
mean source redshift of $0.9$. In this plot the error bars are the
measurement errors and do not include the cosmic variance (see BvWM02).
The dot-dot-dashed line represents the signal corrected from the
residual systematics (the $B$-mode contribution to the two-point
correlation function subtracted from the $B$ mode contribution as
discussed in Pen et al. 2002), while the solid line shows the total
($E+B$) measured signal. The closeness of the two curves reveals the
small amount of residual systematics still present in the 2-point
correlation function.  \\ A quality assessment of the 3-point function
measurements can be done by studying the effect of PSF correction. It
is inferred from the star shapes (they ought to be round) following
the standard procedure described initially in Kaiser, Squires \&
Broadhurst (1995). It can be thought as a star shear field that has to
be subtracted off the measured galaxy shapes. One can then measure the
2- and 3-point correlation functions of the star field and compare
them to the PSF corrected data set. This is presented on Figure
\ref{xistars}.  The solid line in the top and bottom plots
respectively show the star two and three point functions, which are
compared to the same quantities measured on the corrected galaxies
(dashed lines and dashed-dot line for the corrected $E$-mode 2-point
function).  The star two-point correlation function is significantly
larger than the galaxy signal, but this is known not to be a problem
for the two-point function: Erben et al. (2001) and Bacon et al.
(2001) have shown that the PSF correction can account for star
anisotropies as large as $10\%$ (the stars r.m.s. anisotropy is $6\%$
in our data) to a precision better than than one percent.  This is
also demonstrated in BvWM02 using cosmic shear simulated data. The
small amplitude of the $B$-mode is another proof of the robustness of the
correction.

BvWM02 have also shown that the correction scheme works for the shear
three-point function. We see that the situation for the three-point
function (Figure \ref{xistars}, bottom panel) is far more interesting
than the two-point function: the correction is smaller (that is in
principle more robust), and the angular dependence of the star three
point function is totally different from the galaxy
three-point function. If our three-point function signal were
dominated by systematics it would likely not fit the expected signal
as we can see from Figure \ref{xidetec}, but would be more similar to
Figure \ref{xistars}, bottom panel (solid line).  Finally we have
checked that cuts in magnitudes do not significantly change the
results. Therefore, they are not  produced by one single class of objects
of the sample. We have thus some  evidences that what is observed is
genuinely cosmological.

\begin{figure}
\begin{tabular}{c}
{\epsfxsize=7.5cm\epsffile{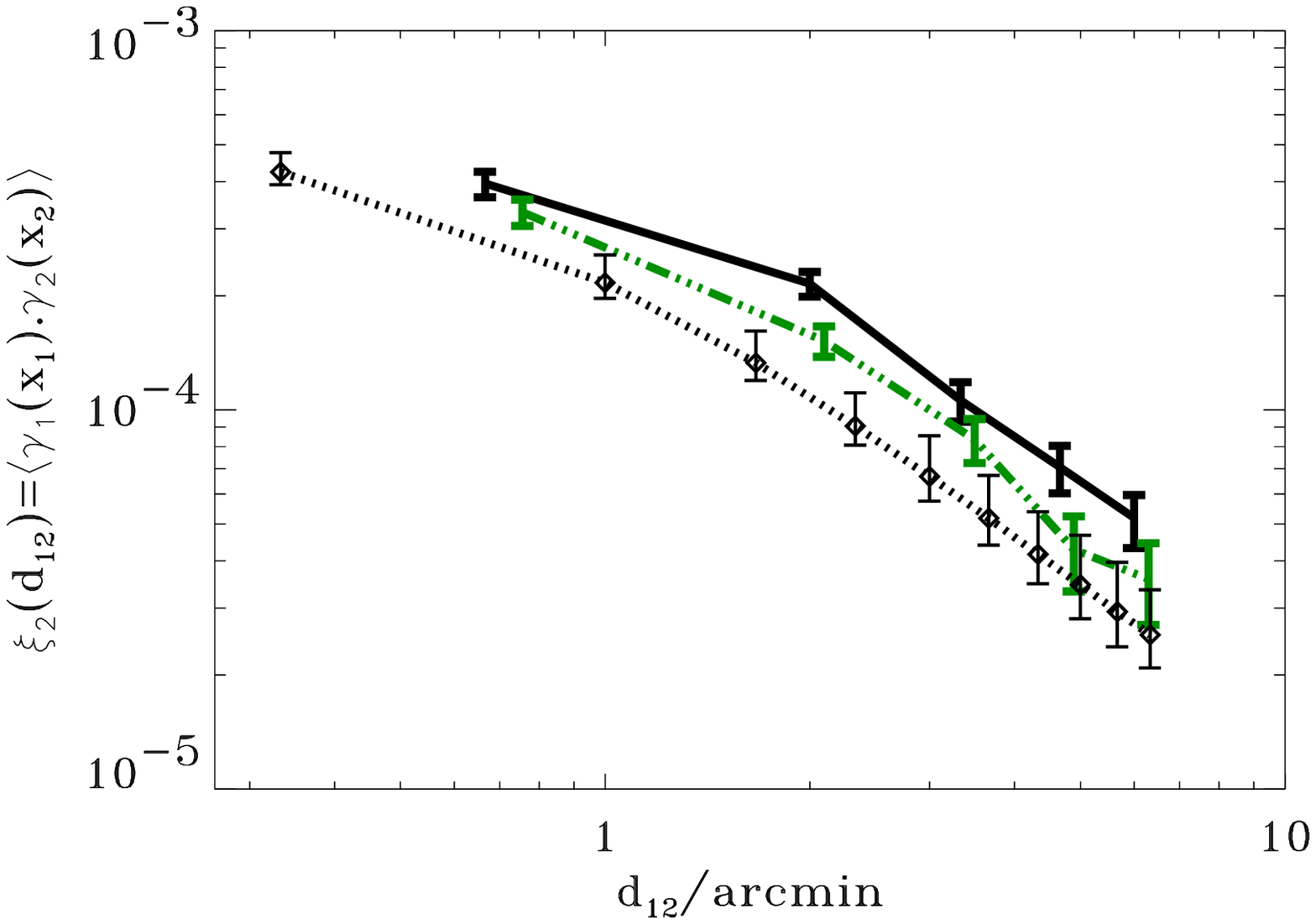}}\\
{\epsfxsize=7.5cm\epsffile{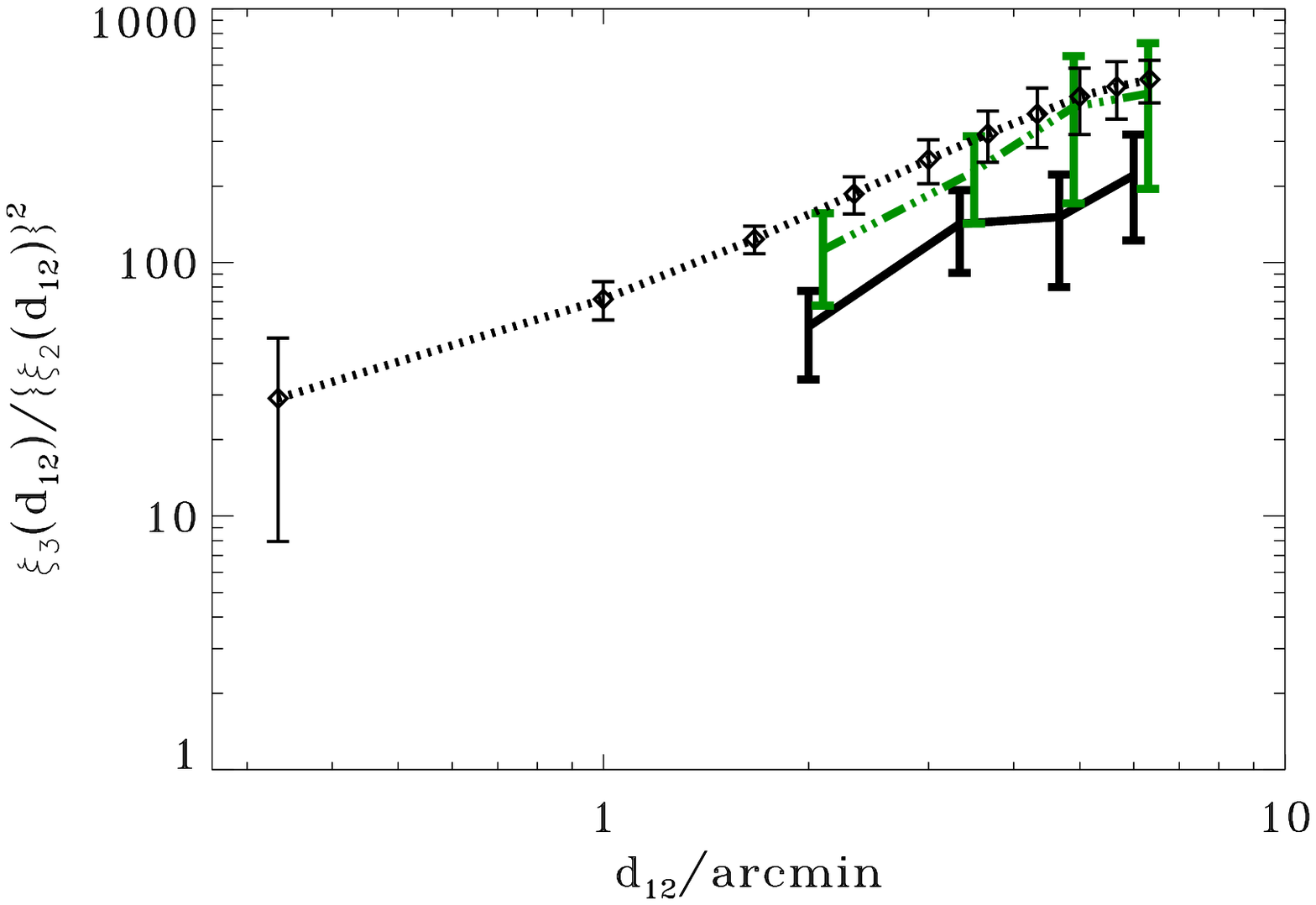}}
\end{tabular}
\caption{Results for the VIRMOS-DESCART survey (similar to Fig. 1)
compared to OCDM results (dotted lines) with cosmic variance
computations.}
\label{xicosvar}
\end{figure}

Another issue, not taken into account in the error bars, is the
so-called cosmic variance, that is the amount by which such a signal
can vary in surveys of finite size. The complete theory of cosmic
variance for such a survey is yet to be done and is likely
  complex (given the topology).  Figure \ref{xicosvar} shows the cosmic
variance obtained from a set of 7 ray-tracing realizations of the open
CDM model (Jain et al. 2000). Note that the error-bars in the
different bins \emph{are correlated}. A rough examination of the
situation shows that while the cosmic variance dominates the error
budget for the two-point function, observational errors and cosmic
variance are of similar amplitude for the three-point function. From
the examination of Fig. \ref{xidetec} $\Lambda$CDM is clearly in
better agreement with the data than the other two models, while
$\tau$CDM is marginally acceptable.

\section{Discussion}
The result shown on Figure \ref{xidetec} is the first detection of
non-Gaussian  features in a cosmic shear survey. The signal is
detected  with a 2.4-$\sigma$ confidence level on 4 independent bins
which gives a 4.9-$\sigma$ global confidence level.  Such a result
opens the route to break the degeneracy between  $\Omega_m$ and
$\sigma_8$ in a way which is independent on assumptions beyond the
solely hypothesis that large-scale structures grows from gravitational
instability of an initial Gaussian field.  We note that the amplitude
of the reduced  3-point correlation function exhibits an angular
dependence which is in agreement with theoretical expectations. It
supports the interpretation of these results as genuine effects of the
gravitational dynamics.

The signal is however still too noisy to provide reliable
information on cosmological parameters. Moreover, several
obstacles have yet to be overcome. It is first important to
understand to which level the measurements are contaminated by
systematics. This can be done through consistency checks yet to be
invented (a statistic which cancels the signal in a non-trivial
way, like the $45$ degrees rotation test for the aperture mass
will have to be found for the three-point function). We have
already checked with the anisotropy of the stars that our signal
is unlikely to be dominated by systematics since it would
otherwise exhibits a totally different angular dependence. In the
next stages, the scientific interpretation of the three-point
function measurements will require a significant improvement over
several issues:

-The knowledge of the redshift distribution of the sources is
crucial for the third order statistics (see BvWM97);

-The source clustering effect might bias the measurement is a
significant way (Bernardeau 1998, Hamana et al. 2002) if the width
of the source distribution is too large;

-Intrinsic alignment of galaxies have a completely unknown effect
on the non-Gaussian properties of the shear field.

Resolving these issues will require progress from both the
theoretical/simulation side and from the observations, which are
already on their way.

{ \acknowledgements  We thank  the VIRMOS and Terapix teams who got
and processed the VIRMOS-DESCART data, B. Jain and M.
White for the use of their  ray-tracing simulations as well as
T. Hamana for useful comments. This work was
supported by the TMR Network ``Gravitational Lensing: New Constraints
on Cosmology and the Distribution of Dark Matter'' of the EC under
contract No. ERBFMRX-CT97-0172. FB thanks IAP for hospitality.}

\end{document}